# Effective Phononic Crystals for Non-Cartesian Elastic Wave Propagation

Ignacio Arretche[1], Kathryn H. Matlack[1]

[1]Department of Mechanical Science and Engineering, University of Illinois Urbana-Champaign, Urbana, IL, USA

We introduce the concept of effective phononic crystals, which combine periodicity with varying isotropic material properties to force periodic coefficients in the elastic equations of motion in a non-Cartesian basis. Periodic coefficients allow for band structure calculation using Bloch theorem. Using the band structure, we demonstrate band gaps and topologically protected interface modes can be obtained in cylindrically propagating waves. Through effective phononic crystals, we show how behaviors of Cartesian phononic crystals can be realized in regions close to sources, where near field effects are non-negligible.

Phononic crystals (PC) and metamaterials have shown great promise when it comes to acoustic and elastic wave propagation control. For example, band gaps can be tailored through geometry [1,2] and tuned through external stimuli [3–5] to prevent select frequency bands from propagating. Through band gap formation and tailored anisotropy, more complex phenomena such as negative refraction can be realized [6,7], which can break the diffraction limit [8] resulting in enhanced imaging. Topologically protected states, stemming from quantum physics, have also been realized in PCs [9–12], which provide robust wave guiding and protection against backscattering.

Despite novel properties of PCs, most studies assume Cartesian plane waves, limiting their application to plane wave excitation or in the far field of sources. The main reason for such assumption is that a plane wave propagating in a medium with invariance to Cartesian translations is described by equations of motion that have periodic coefficients. Their solution satisfies the Bloch theorem [13] and thus the analysis of the infinite system can be reduced to obtaining the band structure of a single unit cell. This is not the case in the near field of a point source, where waves propagate cylindrically in 2-D, or spherically in 3-D. Axisymmetric/cylindrical wave propagation is usually described in terms of Bessel functions that are solutions to differential equations where first and second derivatives are multiplied by the independent variable. For this reason, unlike in the Cartesian basis, a medium with radial translational invariance does not yield equations of motion with periodic coefficients. Thus, the Bloch theorem is not applicable, and properties of PCs based on this type of analysis are not valid in this case.

Still, radially-periodic media has been studied using alternative approaches to calculate approximate band structures, such as using a radial dependent Floquet propagator [14] or by assuming a sufficiently large radius to approximate plane waves [15,16]. Other studies avoid Bloch analysis and study only finite structures with alternating homogeneous rings [17–19]. However, their behavior depends on where the finite system is truncated, and the band structure cannot be calculated.

However, band structure calculation is of great importance in PCs: a major contribution to the exponential growth of the field is the unification of wave phenomena in condensed matter physics, electromagnetism and classical mechanics through band structure analysis; e.g. topologically protected edge wave propagation was first developed for quantum systems [20] and super-resolution was first developed in photonics [8]. Band structure analysis in the polar basis has been done in radial wave crystals [21], which contain heterogeneous media with anisotropic mass density that force periodic coefficients in the scalar wave equation. These materials have been shown to exhibit Fabry-Perot-like resonances [21] and source position detection capabilities [22] in acoustic and electromagnetic anisotropic systems. However, even though anisotropic mass density is possible in radially periodic structures [20], it significantly complicates physical realization. In fact, radial wave crystals have only been physically realized in their electromagnetic version [22]. Further, this approach is limited to acoustic waves in fluids and electromagnetic waves.

In this letter, we extend the concept of radial wave crystals to *elastic* waves, by combining periodicity with radially varying *isotropic* material properties, which we term *effective phononic crystals* (EPC). By explicitly choosing how the material properties depend on radius, we enforce the elastodynamic equation in radial coordinates to contain displacements that satisfy the Bloch theorem, while avoiding the need for anisotropic density. We will show how this approach enables PC properties, such as band gaps and topological edge modes, to occur close to sources where near field effects are significant. We demonstrate how to generally realize PC properties in non-Cartesian systems for elastic waves using radial axisymmetric torsional 2-D waves, which are relevant for e.g. rotating machinery [23] and liquid sensing [24]. Still, our approach can be applied to other polarizations, wave propagation directions, and dimensions. For example, a combination of a radially dependent elastic foundation (Eq. (S5)) and radially dependent material properties (Eq. (S6) and (S7)) will results in periodic coefficients for radially propagating waves (see Supplemental Material (SM) [25]). We specifically show that EPCs allow for band gaps and topological interface modes in the near field.

The equation of motion and constitutive law for radially propagating torsional waves in a 2-D heterogeneous isotropic medium, assuming an axisymmetric displacement field, can be reduced to [26]:

$$\begin{cases} \dfrac{\partial \sigma_{r\theta}(r,t)}{\partial r} + \dfrac{2}{r}\sigma_{r\theta}(r,t) + F_\theta = \rho(r)\dfrac{\partial^2 u_\theta(r,t)}{\partial t^2} \\ \sigma_{r\theta}(r,t) = \mu(r)\left(\dfrac{\partial u_\theta(r,t)}{\partial r} - \dfrac{u_\theta(r,t)}{r}\right) \end{cases} \quad (1)$$

where $\mu$ is the shear modulus, $u_\theta$ is the tangential displacement, $\rho$ is the material density and $F_\theta$ is a tangential body force (see SM for full derivation [25]). Under free vibration and applying separation of space and time (i.e. $u_\theta(r,t) = U_\theta(r)f(t)$) the spatial equation of motion is:

$$\mu(r)\dfrac{\partial^2 U_\theta(r)}{\partial r^2} + \left(\dfrac{\partial \mu(r)}{\partial r} + \dfrac{\mu(r)}{r}\right)\left(\dfrac{\partial U_\theta(r)}{\partial r} - \dfrac{U_\theta(r)}{r}\right) = -\omega^2 \rho(r) U_\theta(r) \quad (2)$$

In Eq. (2), imposing material properties $\mu(r)$ and $\rho(r)$ to be invariant to radial translations of the form $r = r + na$, where $n$ is an integer, will not result in an ordinary differential equation with periodic coefficients. Thus, we cannot apply Bloch theorem nor calculate band structure. We instead define a set of isotropic material properties that are not invariant to translation but do enforce periodic coefficients in Eq. (2). This type of material becomes a phononic material not from *geometrical periodicity*, i.e. invariance of geometry to lattice constant translations in the direction of the basis vectors, but because their material properties describe an effectively periodic system. Thus, we define them as *effective phononic crystals* (EPC). Note that Cartesian PCs are a subset of EPCs that also have geometric periodicity. Like Cartesian PCs, there is not a unique way of designing an EPC. Here, we present two possible designs, one targeting effective periodicity in terms of *angular displacements* (EPC1) and the other in terms of *tangential displacements* (EPC2).

One way of realizing an EPC is to rewrite Eq. (2) in terms of angular displacements ($\Phi(r) = U_\theta(r)/r$):

$$\frac{\partial}{\partial r}\left(\mu(r) r^3 \frac{\partial \Phi(r)}{\partial r}\right) = -\omega^2 \rho(r) r^3 \Phi(r) \tag{3}$$

We now define material properties in a piecewise fashion such that,

$$\mu(r) = \begin{cases} M_1/r^3 & r \in D_1 \\ M_2/r^3 & r \in D_2 \end{cases} \tag{4}$$

$$\rho(r) = \begin{cases} P_1/r^3 & r \in D_1 \\ P_2/r^3 & r \in D_2 \end{cases} \tag{5}$$

where $D_1 = \{r \in \mathrm{IR} > 0 \,|\, r_0 + (n-1)a \leq r < r_0 + na - a_2\}$, $D_2 = \{r \in \mathrm{IR} > 0 \,|\, r_0 + na - a_2 \leq r < r_0 + na\}$ (Fig. 1(a)), $r_0$ is the internal radius of the EPC, $a = a_1 + a_2$ is the periodicity constant, $a_i$ is the thickness of heterogeneous layer $i$, $n$ is an integer and $M_i$, $P_i$ are constant in each layer (Fig. 1(b)). Essentially, the system consists of alternating ring layers with heterogeneous isotropic material properties. We refer to this EPC as EPC1. With these material properties, Eq. (3) is reduced to a second order ordinary differential equation with periodic coefficients,

$$\frac{\partial^2 \Phi_i(r)}{\partial r^2} + \frac{P_i}{M_i} \omega^2 \Phi_i(r) = 0 \tag{6}$$

Note that: (i) the material properties are not geometrically periodic, and (ii) the periodic coefficients are obtained in terms of angular displacements, and thus, this material will behave as an EPC in terms of angular displacements. We can now apply Bloch theorem and moment and displacement continuity between layers to obtain the band structure of the EPC [13]:

$$\cos(Ka) = \cos(k_1 a_1) \cos(k_2 a_2) - \frac{1+z^2}{2z} \sin(k_1 a_1) \sin(k_2 a_2) \tag{7}$$

where $k_i^2 = \omega^2 P_i / M_i$, $z^2 = M_2 P_2 / M_1 P_1$ and $K$ is the Bloch wave number in the radial direction (see SM [25] for full derivation). This dispersion relation is analogous to that of an equivalent bilayer Cartesian PC with elastic moduli $M_i$, density $P_i$, and layer thickness $a_i$ [27]. Thus, we expect properties of Cartesian bilayer system to apply to this EPC.

A second way to obtain an EPC is to define the shear modulus in such a way that the second term in Eq. (2) is zero for all $U_\theta(r)$, and selectively define density. This way, we recover an analogous equation to Eq. (6) but in terms of tangential displacements. We do this by setting,

$$\mu(r) = \begin{cases} M_1/r & r \in D_1 \\ M_2/r & r \in D_2 \end{cases} \tag{8}$$

$$\rho(r) = \begin{cases} P_1/r & r \in D_1 \\ P_2/r & r \in D_2 \end{cases} \quad (9)$$

We refer to this EPC as EPC2 (Fig. 1(c)), where the tangential displacements follow an ordinary differential equation with periodic coefficients (i.e. we can apply Bloch theorem to the tangential displacement equation). However, due to the dependence of stress on the first derivative of the tangential displacement (Eq. (1)), the band structure is not equal to Eq. (7) but asymptotically approaches it as frequency increases. In fact, the transmission of EPC1 is quite close to the transmission of a Cartesian PC well below the Bragg frequency (Fig. 2(e)). The reader is referred to the SM [25] for more details on the dispersion relation calculations.

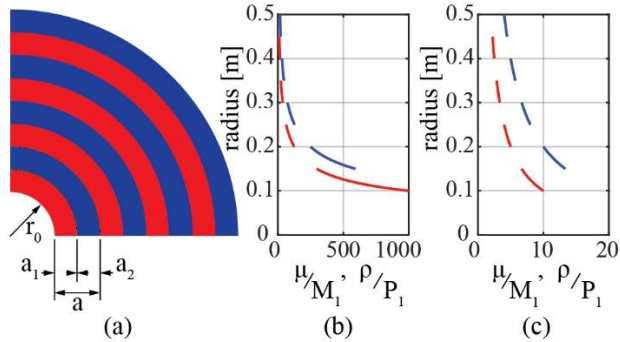

**FIG 1.** (a) Quarter section of the effective phononic crystal for torsional wave propagation. (b) Radial dependence of material properties for EPC1 (Eq. (4)-(5)). (c) Radial dependence of material properties for EPC2 (Eq. (8)-(9)). The EPCs shown here have $r_0/a = 1$.

To validate our approach, we calculate transmission of radial torsional waves using finite element analysis for finite 4-unit cell EPC1 and EPC2 (Figs. 2(d) and 2(e)). As a comparison, we also calculate transmission of a material with homogeneous alternating ring layers with shear modulus and density of layer $i$ equal to $M_i$ and $P_i$, respectively (Fig. 2(c)). This material is geometrically periodic but not effectively periodic (Eq. 2). In all three systems, we impose a harmonic tangential displacement on the inner boundary and traction free boundary conditions on the outer boundary. Since EPC1 has effective periodicity in terms of angular displacement, transmission is calculated as the ratio of outer boundary to inner boundary angular displacement. For EPC2 transmission is calculated based on tangential displacements. As a benchmark, we also calculate transmission of plane Cartesian waves propagating in the equivalent Cartesian PC using the transfer matrix method.

The transmission of the homogeneous layered system is strongly dependent on $r_0/a$ (Fig 2(c)), and deviates in amplitude and frequency from the equivalent Cartesian system at small $r_0/a$. This means that transmission also depends on where we truncate this material. This demonstrates that pure radial tessellations do not result in effective periodicity. For small $r_0/a$, the modes shift toward higher frequencies, shifting the transmission reduction region (usually associated with a band gap) toward higher frequencies (Fig. 2(c)-red curve). This can be explained by considering the solution of an outward-propagating radial torsional wave propagating in a semi-infinite homogeneous medium (see SM [25]). The phase velocity of this wave is inversely proportional to radius and asymptotically approaches the bulk material shear wave speed as radius increases (see Fig S2(c) in [25]). The phase velocity of the first few rings is higher and thus the overall frequency shifts. As the ratio between internal radius and lattice constant increases, the transmission approaches that of a Cartesian PC. Essentially, in the far field, radially propagating waves approximate plane wave fronts.

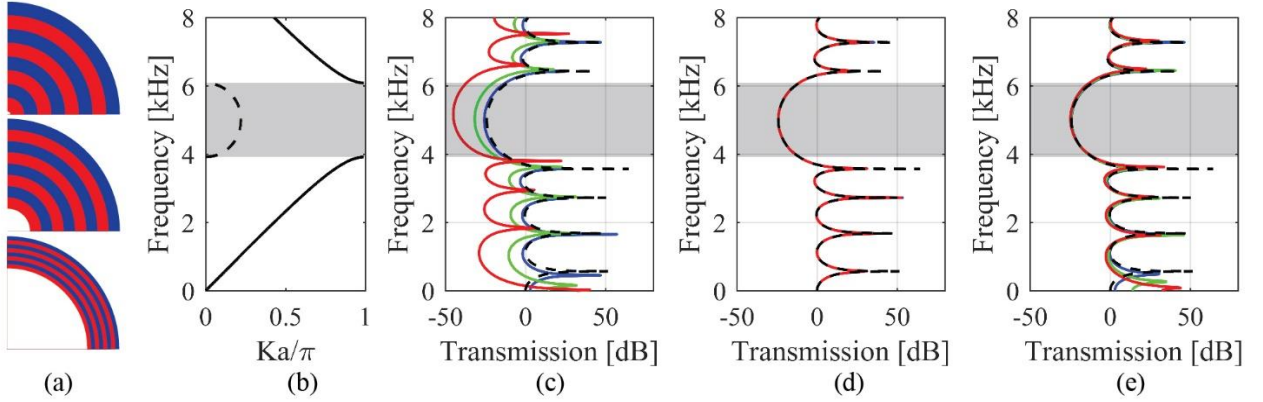

**FIG 2**. (a) Quarter section of the EPC for $r_0/a = 0.1$ (top), $r_0/a = 1$ (middle), $r_0/a = 10$ (bottom). Images do not have the same scale. (b) Dispersion relation for the EPCs (Eq. 7) for $a_1 = a_2 = 0.05m$ (dashed lines show imaginary Bloch wave number). Transmission curves for: (c) radially alternating homogeneous rings, (d) EPC1, and (e) EPC2, each plotted for $r_0/a = 0.1$ (red), $r_0/a = 1$ (green), $r_0/a = 10$ (blue), and compared to the equivalent Cartesian PC (black-dashed).

For both EPCs, we observe very good agreement between transmission reduction and the band gap predicted from the dispersion relation (Fig. 2(a) and 2(d)-2(e)). For EPC1, the dispersion relation is exact and equivalent to that of a Cartesian PC. Note that this specific material behaves periodic only to angular displacements, thus Bloch wave homogenization in the long wavelength region can only be applied for this propagation direction and polarization. This will result in an effective density and shear modulus in $r\theta$ plane that are equivalent to those of the Cartesian PC. Since the system does not behave as a PC in the other directions/polarizations, other methods of homogenization (i.e. an elastostatic approach [28]) must be used to calculate the remaining effective material properties. Like other 1D periodic media, these EPCs will generally have anisotropic effective material properties. The transmission curves are independent of unit cell truncation for all frequencies and exactly correspond to the Cartesian system. Note that even though the EPC is defined for a semi-infinite media (i.e. $r \geq 0$), the results are independent of unit cell truncation even when $r_0/a < 1$. This means that even though we cannot physically add a unit towards smaller radii, the system still behaves as infinitely periodic.

For EPC2, the band structure (Fig. 2(a)) accurately characterizes the EPC above a certain frequency (thus Bloch wave homogenization in the long-wavelength region is not possible in this case), which is well below the first Bragg frequency. Above this frequency, the transmission through EPC2 is independent of unit cell truncation and approximates that of the Cartesian PC (Fig. 2(e)). It is crucial that the band structure is accurate well below the band gap frequencies since many interesting properties of PCs arise from band gap formation. The response of this EPC is a dynamic effect: as we approach a quasi-static condition (frequency approaches zero), a tangential displacement in the inner boundary does not result in an equal tangential displacement in the outer boundary.

The benefit of band gaps in PCs is that mechanical energy can be effectively transmitted over a band of frequencies, while another band of frequencies is effectively attenuated. In this sense, since the EPCs (Fig. 2(d) and 2(e)) exhibit this behavior, they are clearly superior to the homogeneous layered material (Fig. 2(c)), which does not effectively transmit energy in propagating bands. However, in terms of absolute vibration mitigation over all frequencies, the homogeneous layered material actually provides the largest vibration mitigation, due to diffraction as the wave propagates outward. However, this will not be true if the wave were to propagate inward, as in the case of vibrations from the teeth of a gear to its center shaft. In the latter case, the wave front expansion would *increase* vibration amplitudes resulting in higher transmission. Instead, the EPCs mimic a Cartesian system and thus diffraction is effectively compensated by the prescribed material properties, irrespective of the wave propagation direction. For inward propagation, this will result in lower vibration transmission inside the band gap compared to the homogeneous layered material.

The EPCs presented here are indeed not unique but represent a subset of possible ways to overcome issues with non-plane and near field vibrations. Through EPCs, we can embed behaviors of Cartesian PCs and metamaterials in non-Cartesian waves by redefining the material properties. To highlight this last point, we next demonstrate topologically protected modes in the EPCs.

Topological concepts in mechanical systems arise from a correlation between bulk electron bands in a crystal and bulk vibration bands in periodic lattices [29]. Both electrons and phonons in periodic media can be characterized with Bloch wave solutions, and thus topological quantities based on these can be applied to both domains. However, such behaviors cannot be embedded in material with radial geometric periodicity, since these structures are not described by Bloch wave functions. Instead, we use EPCs that are effectively periodic, to enable topological properties to be applied to radially propagating waves.

We target a topological interface mode since the EPCs analyzed here consider 1-D wave propagation. It is well known that a topologically protected mode is generated at the interface of two phononic crystals with different topological properties. This has been well developed in 1-D plane waves for acoustic [9,30] and elastic waves [10,12,31] but to the authors' knowledge it has not yet been shown in other basis systems.

To obtain a topological interface mode in the cylindrical basis, we design an EPC superlattice made of two EPC sublattices that follow the formulation of Eq. (4)-(5) and (8)-(9) (Fig. 3(a)) and $r_0/a = 0.1$. Sublattice A has $a_1 = 1/3a$ and $a_2 = 2/3a$, and sublattice B has $a_1 = 2/3a$ and $a_2 = 1/3a$, where $a = 0.1m$. Both sublattices have the same band structure (Fig. 3(b)) but different topological invariants in their second band: sublattice A has a second band Zak phase of 0 and sublattice B a second band Zak phase of $\pi$ (see SM [25]). Because of the difference in topology, a topological interface mode between sublattices arises inside the second band gap (Fig 3(c)). We calculate transmission from inner ring to the interface and compare superlattices made of EPC1, EPC2, homogeneous layer rings, and the equivalent Cartesian superlattice (Fig. 3(c)).

Transmission results show a mode inside the second band gap for both EPC1, EPC2 and the homogeneous layered rings, suggestive of an interface mode. However, we will show that only EPC1 and EPC2 support an interface mode. The mode corresponding to the homogeneous layered system is shifted to a higher frequency for the same reason as the band gap shift (Fig. 2(c)). To characterize how localized this mode at the interface, we run an eigenfrequency analysis and plot the modal displacement of the interface mode normalized by the maximum value (Fig. 3(d)). While there is some localization in all systems, the homogeneous system clearly does not show the characteristic behavior of a topological interface mode: there is no increase in oscillating displacements from source boundary ($\bar{x}=0$) to the interface ($\bar{x}=0.5$). This is particularly evident close to the inner boundary where the effect of diffraction is more significant. In fact, close to the source, modal displacements are about 5 times larger than those of EPC1, EPC2 and the equivalent Cartesian PC.

For EPC1, the interface mode is clearly present and there is a localization at the interface compared to the source boundary. The modal displacements of EPC1 match those of the Cartesian PC, because the dispersion relation is exactly the same as the equivalent Cartesian PC. The interface mode is also present in EPC2. Even though the dispersion relation of EPC2 is an approximation of the Cartesian PC, the band structure at these high frequencies accurately captures the behavior of EPC2, and differences in modal displacements of EPC2 compared to the Cartesian PC are almost negligible (Fig. 3(d) - inset). Through EPCs, we demonstrate for the first time a topological interface mode with a cylindrical wave (Fig. 3(e)).

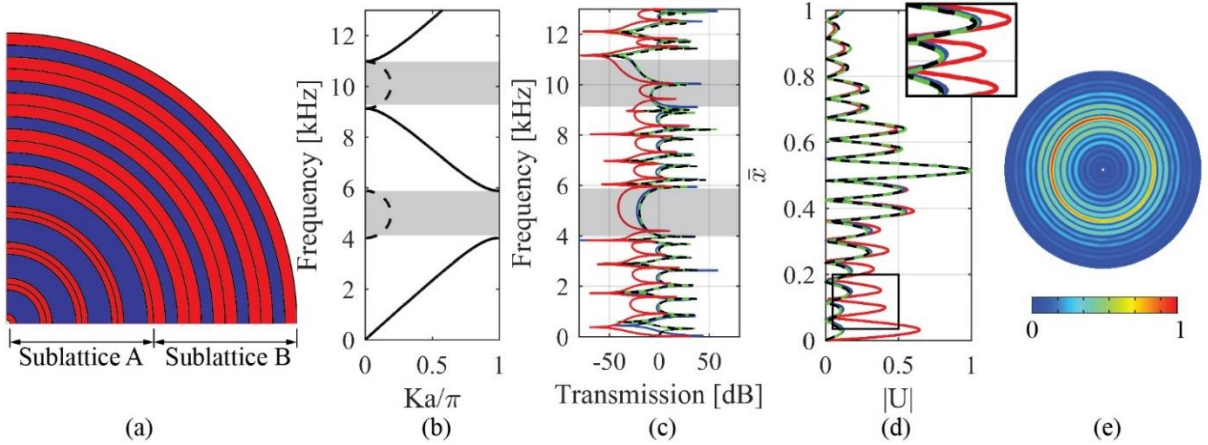

**FIG 3.** (a) Quarter section of the superlattice for $r_0/a = 0.1$ (b) Dispersion relation for the EPCs sublattices (dashed lines show imaginary Bloch wave number and shaded regions band gap frequencies). (c) Transmission from inner boundary to interface of the superlattice for homogeneous layers (red), EPC1 (green), EPC2 (blue), equivalent Cartesian PC (black-dashed). (d) Magnitude of modal displacement at interface mode frequency for homogeneous layers (red), EPC1 (green), EPC2 (blue), equivalent Cartesian PC (black-dashed) in terms of normalized spatial coordinate ($\bar{x} = (r - r_0)/(r_{max} - r_0)$ for radial PCs and $\bar{x} = x/x_{max}$ for Cartesian PC). The inset shows zoom-in view of modal displacements. (e) Normalized modal displacements of the EPC1 at the interface mode frequency.

In this letter, we introduce the concept of effective phononic crystals, which combine radially-dependent isotropic material properties with geometric periodicity, such that the Bloch theorem can be applied to non-Cartesian elastic waves in non-periodic media. Further, our approach enables non-Cartesian elastic waves to exhibit PC properties, such as band gaps and topologically edge modes. By analyzing finite EPCs, we show that the calculated band structure of the EPCs using Bloch theorem is valid and accurately approximates that of the Cartesian PC. Using this approach, we demonstrate topological interface modes for cylindrically propagating waves, which do not occur in homogeneous radially periodic systems.

The EPCs require isotropic mass density, thus simplifying physical realization compared to anisotropic mass density [21]. Even though continuously changing material properties could be challenging to physically realize, we show that discretization of material properties within each layer yields a similar approximate behavior (see Fig. S3(a) and S3(b) in [25]). In this way, each layer could consist of one or more layers of different lattice geometries, such as stretch-dominated lattices that exhibit the required linear dependence of effective modulus on density [32,33], which could be fabricated with techniques such as voxel-based additive manufacturing [34]. EPCs may allow for improved focusing, subwavelength band gaps, negative refraction properties, topological states and other novel wave propagation control, that occur close to sources where near filed effects limit the application of conventional PCs.

The authors acknowledge support from a University Research Program from Ford Motor Company.**REFERENCES**

[1]  I. Arretche and K. H. Matlack, Front. Mater. **5**, 68 (2018).

[2]  K. H. Matlack, A. Bauhofer, S. Krödel, A. Palermo, and C. Daraio, Proc. Natl. Acad. Sci. **113**, 8386 (2016).

[3]  C. Nimmagadda and K. H. Matlack, J. Sound Vib. **439**, 29 (2019).

[4]  P. Wang, F. Casadei, S. Shan, J. C. Weaver, and K. Bertoldi, Phys. Rev. Lett. **113**, 014301 (2014).

[5]  C. D. Pierce, C. L. Willey, V. W. Chen, J. O. Hardin, J. D. Berrigan, A. T. Juhl, and K. H. Matlack, Smart Mater. Struct. **29**, 065004 (2020).

[6]  K. H. Matlack, M. Serra Garcia, M. Serra-Garcia, A. Palermo, S. D. Huber, and C. Daraio, Nat. Mater. **17**, 323 (2018).

[7]  S. Yang, J. H. Page, Z. Liu, M. L. Cowan, C. T. Chan, and P. Sheng, Phys. Rev. Lett. **93**, 024301 (2004).

[8]  J. B. Pendry, Phys. Rev. Lett. **85**, 3966 (2000).

[9]  M. Xiao, G. Ma, Z. Yang, P. Sheng, Z. Q. Zhang, and C. T. Chan, Nat. Phys. **11**, 240 (2015).

[10]  J. Yin, M. Ruzzene, J. Wen, D. Yu, L. Cai, and L. Yue, Sci. Rep. **8**, 6806 (2018).

[11]  Z. Yang, F. Gao, X. Shi, X. Lin, Z. Gao, Y. Chong, and B. Zhang, Phys. Rev. Lett. **114**, 114301 (2015).

[12]  R. Chaunsali, E. Kim, A. Thakkar, P. G. Kevrekidis, and J. Yang, Phys. Rev. Lett. **119**, 024301 (2017).

[13]  Brillouin, Léon, *Wave Propagation in Periodic Structures; Electric Filters and Crystal Lattices,* 1st ed. (McGraw-Hill, New York, 1946).

[14]  A. Hvatov and S. Sorokin, J. Sound Vib. **414**, 15 (2018).

[15]  Z. Xu, F. Wu, and Z. Guo, Phys. Lett. A **376**, 2256 (2012).

[16]  T. Ma, T. Chen, X. Wang, Y. Li, and P. Wang, J. Appl. Phys **116**, 104505 (2014).

[17]  S. Haisheng, D. Liqiang, L. Shidan, L. Wei, L. Shaogang, W. Weiyuan, S. Dongyan, and Z. Dan, J. Phys. D. Appl. Phys. **47**, 295501 (2014).

[18]  P. Yeh, A. Yariv, and E. Maron, J Opt Soc Am **68**, 1196 (1978).

[19]  H. Shu, L. Zhao, X. Shi, W. Liu, D. Shi, and F. Kong, J. Appl. Phys **118**, 184904 (2015).

[20]  R. K. Pal, J. Vila, and M. Ruzzene, in *Adv. Cryst. Elastic Metamaterials, Part 2*, edited by M. I. Hussein (Elsevier, 2019), pp. 147–181.

[21]  D. Torrent and J. Sánchez-Dehesa, Phys. Rev. Lett. **103**, 064301 (2009).

[22]  J. Carbonell, A. Díaz-Rubio, D. Torrent, F. Cervera, M. A. Kirleis, A. Piqué, and J. Sánchez-Dehesa, Sci. Rep. **2**, 558 (2012).8

# Supplemental Material

## DERIVATION OF EQUATIONS OF MOTION

From conservation of linear momentum, the stress equations in cylindrical coordinates are [1]:

$$\begin{cases} \dfrac{\partial \sigma_{rr}}{\partial r} + \dfrac{1}{r}\dfrac{\partial \sigma_{r\theta}}{\partial \theta} + \dfrac{\partial \sigma_{zr}}{\partial z} + \dfrac{1}{r}(\sigma_{rr} - \sigma_{\theta\theta}) + F_r = \rho \dfrac{\partial^2 u_r}{\partial t^2} \\ \dfrac{\partial \sigma_{r\theta}}{\partial r} + \dfrac{1}{r}\dfrac{\partial \sigma_{\theta\theta}}{\partial \theta} + \dfrac{\partial \sigma_{z\theta}}{\partial z} + \dfrac{2}{r}\sigma_{r\theta} + F_\theta = \rho \dfrac{\partial^2 u_\theta}{\partial t^2} \\ \dfrac{\partial \sigma_{rz}}{\partial r} + \dfrac{1}{r}\dfrac{\partial \sigma_{\theta z}}{\partial \theta} + \dfrac{\partial \sigma_{zz}}{\partial z} + \dfrac{1}{r}\sigma_{rz} + F_z = \rho \dfrac{\partial^2 u_z}{\partial t^2} \end{cases} \quad (S10)$$

Assuming infinitesimal strains, the kinematic equations in cylindrical coordinates are [1]:

$$\begin{cases} \varepsilon_{rr} = \dfrac{\partial u_r}{\partial r} \\ \varepsilon_{\theta\theta} = \dfrac{1}{r}\dfrac{\partial u_\theta}{\partial \theta} + \dfrac{u_r}{r} \\ \varepsilon_{zz} = \dfrac{\partial u_z}{\partial z} \\ \varepsilon_{r\theta} = \dfrac{1}{2}\left(\dfrac{1}{r}\dfrac{\partial u_r}{\partial \theta} + \dfrac{\partial u_\theta}{\partial r} - \dfrac{u_\theta}{r}\right) \\ \varepsilon_{\theta z} = \dfrac{1}{2}\left(\dfrac{\partial u_\theta}{\partial z} + \dfrac{1}{r}\dfrac{\partial u_z}{\partial \theta}\right) \\ \varepsilon_{zr} = \dfrac{1}{2}\left(\dfrac{\partial u_r}{\partial z} + \dfrac{\partial u_z}{\partial r}\right) \end{cases} \quad (S11)$$

The constitutive law for an isotropic material can be expressed as:

$$\sigma_{ij} = 2\mu\varepsilon_{ij} + \lambda\varepsilon_{kk}\delta_{ij} \quad (S12)$$

Using Eq. (S1-S3) and assuming plain strain torsional axisymmetric waves (i.e. $\underline{u}(r,\theta,z) = (0, u_\theta(r,t), 0)$) and axisymmetric heterogeneous material properties (i.e. $\mu = \mu(r)$, $\lambda = \lambda(r)$, $\rho = \rho(r)$) the equations of motion can be reduced to Eq. (1).

## EPC FOR RADIALLY POLARIZED WAVES

Using Eq. (S1-S3) and assuming plain strain radial axisymmetric waves (i.e. $\underline{u}(r,\theta,z) = (u_r(r,t), 0, 0)$) and axisymmetric heterogeneous isotropic material properties (i.e. $\mu = \mu(r)$, $\lambda = \lambda(r)$, $\rho = \rho(r)$) the equations of motion can be reduced to:

$$\left(\frac{\partial(\lambda(r)+2\mu(r))}{\partial r}+\frac{\lambda(r)+2\mu(r)}{r}\right)\frac{\partial u_r}{\partial r}-\frac{\lambda(r)+2\mu(r)}{r^2}u_r+(\lambda(r)+2\mu(r))\frac{\partial^2 u_r}{\partial r^2}$$
$$+F_r(r)=\rho\frac{\partial^2 u_r}{\partial t^2} \tag{S13}$$

To obtain an ordinary differential equation with periodic coefficients, we first introduce a radially dependent forcing:

$$F_r(r)=\frac{\lambda(r)+2\mu(r)}{r^2}u_r \tag{S14}$$

We can then obtain an equation of motion with periodic coefficients by setting material properties to be:

$$\lambda(r)+2\mu(r)=\begin{cases} M_1/r & r\in D_1 \\ M_2/r & r\in D_2 \end{cases} \tag{S15}$$

$$\rho(r)=\begin{cases} P_1/r & r\in D_1 \\ P_2/r & r\in D_2 \end{cases} \tag{S16}$$

where $D_1=\{r\in \mathrm{IR}>0\,|\,r_0+(n-1)a\leq r<r_0+na-a_2\}$, $D_2=\{r\in \mathrm{IR}>0\,|\,r_0+na-a_2\leq r<r_0+na\}$ (Fig. 1(a)), $r_0$ is the internal radius of the EPC, $a=a_1+a_2$ is the periodicity constant, $a_i$ is the thickness of heterogeneous layer $i$, $n$ is an integer and $M_i$, $P_i$ are constant in each layer (Fig. 1(b)). Note that the forcing now becomes $(M_i/r^3)u_r$ inside layer $i$. This corresponds to the forcing coming from a linear elastic foundation with stiffness, $k=-M_i/r^3$. Although the stiffness is negative, we could achieve a negative dynamic stiffness using a metamaterial foundation. Dispersion relation of this material can be calculated is the same way as that of the EPC2. Note that this analysis assumes isotropic properties inside each layer (Eq. (S3)). Using an anisotropic constitutive law may avoid the need for the elastic foundation.

## DISPERSION CALCULATION FOR TORSIONAL WAVES

For EPC1, we write the equations in terms of angular displacements (Eq. (3)) and apply material properties according to Eq. (4)-(5) to obtain Eq. (6). In layer $i$, Eq. (6) has solutions of the form,

$$\Phi_i(r)=A_i e^{jk_i r}+B_i e^{-jk_i r} \tag{S17}$$

where $\Phi_i(r)$ are angular displacements in layer $i$, and $A_i$, $B_i$ are constants. From Eq. (1), and assuming separation of space and time, we can write the stress in terms of angular displacements as,

$$\sigma_{r\theta_i}(r)=\frac{M_i}{r^3}\frac{\partial \Phi_i(r)}{\partial r} \tag{S18}$$

We now take an arbitrary unit cell centered at radial coordinate $r_n$ (Fig. S1) and impose continuity of displacements and moments at layer interfaces (layers are numbered according to Fig. S1(b)):

$$\begin{cases} \Phi_1(r_n) = \Phi_2(r_n) \\ T_1(r_n) = T_2(r_n) \\ \Phi_2(r_n + a_2) = \Phi_3(r_n + a_2) \\ T_2(r_n + a_2) = T_3(r_n + a_2) \end{cases} \quad (S19)$$

where $T_i(r) = 2\pi r h \sigma_{r\theta_i}(r)$ is the torsional moment in layer $i$ and $h$ is the thickness of the EPC in the axial direction. Since Eq. (6) is an ordinary differential equation with periodic coefficients, we can apply Bloch theorem [2],

$$\Phi_1(r_n - a_1) = \Phi_3(r_n + a_2) e^{jKa} \quad (S20)$$

Using Eq. (S4)-(S7) we obtain the following algebraic system of equations,

$$\begin{cases} A_1 e^{jk_1 r_n} + B_1 e^{-jk_1 r_n} - A_2 e^{jk_2 r_n} - B_2 e^{-jk_2 r_n} = 0 \\ A_1 e^{jk_1 r_n} - B_1 e^{-jk_1 r_n} - z\left(A_2 e^{jk_2 r_n} - B_2 e^{-jk_2 r_n}\right) = 0 \\ A_1 e^{jk_1(r_n - a_1)} + B_1 e^{-jk_1(r_n - a_1)} - \left(A_2 e^{jk_2(r_n + a_2)} + B_2 e^{-jk_2(r_n + a_2)}\right) e^{jKa} = 0 \\ A_1 e^{jk_1(r_n - a_1)} - B_1 e^{-jk_1(r_n - a_1)} - z\left(A_2 e^{jk_2(r_n + a_2)} - B_2 e^{-jk_2(r_n + a_2)}\right) e^{jKa} = 0 \end{cases} \quad (S21)$$

Note from the last equation in Eq. (S8) that the derivative of the angular displacements also comply with a Bloch-Floquet boundary condition and that Eq. (S8) is analogous to those of a Cartesian layered system [3]. We can then solve for non-trivial solutions of the homogenous system of equation to obtain the dispersion relation (Eq. (7)).

For EPC2, we impose material properties according to Eq. (8)-(9) into Eq. (2) to obtain,

$$\frac{\partial^2 U_{\theta_i}(r)}{\partial r^2} + \frac{P_i}{M_i} \omega^2 U_{\theta_i}(r) = 0 \quad (S22)$$

In layer $i$, Eq. (S9) has a solution of the form,

$$U_{\theta_i}(r) = A_i e^{jk_i r} + B_i e^{-jk_i r} \quad (S23)$$

Similar to EPC1, we take an arbitrary unit cell, impose continuity of displacements and moments at the interfaces, and apply Bloch theorem to Eq. (S9).

$$\begin{cases} U_{\theta_1}(r_n) = U_{\theta_2}(r_n) \\ T_1(r_n) = T_2(r_n) \\ U_{\theta_2}(r_n + a_2) = U_{\theta_3}(r_n + a_2) \\ T_2(r_n + a_2) = T_3(r_n + a_2) \end{cases} \quad (S24)$$

and Bloch theorem to Eq. (S9),

$$U_{\theta_1}(r_n - a_1) = U_{\theta_3}(r_n + a_2)e^{jKa} \quad (S25)$$

Using Eq. (1) to relate moments and displacements, Eq. (S10-S12) and algebraic manipulation, we obtain the following algebraic system of equations,

$$\begin{cases} A_1 e^{jk_1 r_n} + B_1 e^{-jk_1 r_n} - A_2 e^{jk_2 r_n} - B_2 e^{-jk_2 r_n} = 0 \\ A_1 e^{jk_1 r_n} - B_1 e^{-jk_1 r_n} - z\left(A_2 e^{jk_2 r_n} - B_2 e^{-jk_2 r_n}\right) = -\sqrt{\frac{P_1}{M_1}} \frac{M_2 - M_1}{j\omega r_n}\left(A_1 e^{jk_1 r_n} + B_1 e^{-jk_1 r_n}\right) \\ A_1 e^{jk_1(r_n - a_1)} + B_1 e^{-jk_1(r_n - a_1)} - \left(A_2 e^{jk_2(r_n + a_2)} + B_2 e^{-jk_2(r_n + a_2)}\right)e^{jKa} = 0 \\ A_1 e^{jk_1(r_n - a_1)} - B_1 e^{-jk_1(r_n - a_1)} - z\left(A_2 e^{jk_2(r_n + a_2)} - B_2 e^{-jk_2(r_n + a_2)}\right)e^{jKa} = \\ \qquad -\sqrt{\frac{P_1}{M_1}} \frac{M_2 - M_1}{j\omega(r_n + a_2)}\left(A_1 e^{jk_1(r_n - a_1)} + B_1 e^{-jk_1(r_n - a_1)}\right) \end{cases} \quad (S26)$$

Similar to EPC1, we can calculate the dispersion for EPC2 by solving for a non-trivial solution. Because of the right hand side of the second and fourth equations on Eq. (S13), the dispersion relation for EPC2 is not independent of position ($r_n$), thus the system does not strictly behave as an infinitely periodic system. However, we observe that these terms become smaller as frequency and radius increases. Thus, Eq. (S13) asymptotically approximates Eq. (S8), and dispersion relation for EPC2 asymptotically approaches that of EPC1 and the Cartesian layered system as frequency and radius increase.

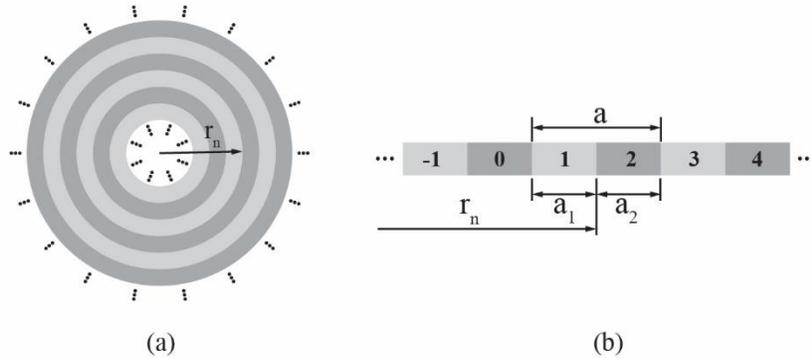

**FIG S1.** (a) Semi-infinite EPC, (b) Cross section of semi-infinite EPC with numbered layers and dimensions

## TORSIONAL WAVES IN A HOMOGENOUS MEDIUM

From Eq. (1), assuming a homogenous medium and a harmonic (i.e. $u_\theta(r,t) = U_\theta(r)e^{-j\omega t}$) we can rewrite the equation of motion as,

$$r^2 \frac{\partial^2 U_\theta(r)}{\partial r^2} + r \frac{\partial U_\theta(r)}{\partial r} + \left(r^2 \beta^2 - 1\right) U_\theta(r) = 0 \quad (S27)$$

where $\beta^2 = \omega^2/c_s^2$ and $c_s$ is the shear wave speed of the homogenous medium. The solution of this equation can be expressed in terms of Hankel function as [1],

$$U_\theta(r) = AH_1^{(1)}(\beta r) + BH_1^{(2)}(\beta r) \tag{S28}$$

where $A$ and $B$ are constants that depend on boundary conditions. By asymptotic representation, it can be shown that the Hankel function of the first kind represents outward-propagating waves and the Hankel function of the second kind inward-propagating waves [1]. Thus, for an outward-propagating wave in a semi-infinite medium (Fig. S2(a)), $B = 0$. Assuming a harmonic excitation with amplitude $u_0$ at $r = r_0$,

$$u_\theta(r,t) = \frac{u_0}{H_1^{(1)}(\beta r_0)} H_1^{(1)}(\beta r) e^{-j\omega t} = |U_\theta(\beta r)| e^{j(\phi(\beta r) - \omega t)} \tag{S29}$$

where $\phi(\beta r) = \angle U_\theta(\beta r)$. The phase velocity ($c_{ph}$) can then be calculated as,

$$c_{ph}(\omega, r) = \frac{\omega}{\partial \phi(\beta r)/\partial r} \tag{S30}$$

Normalized amplitudes and phase velocities for $r_0 = 0.01m$ are plotted as a function of radius for different frequencies in Fig. S2(b) and S2(c), respectively.

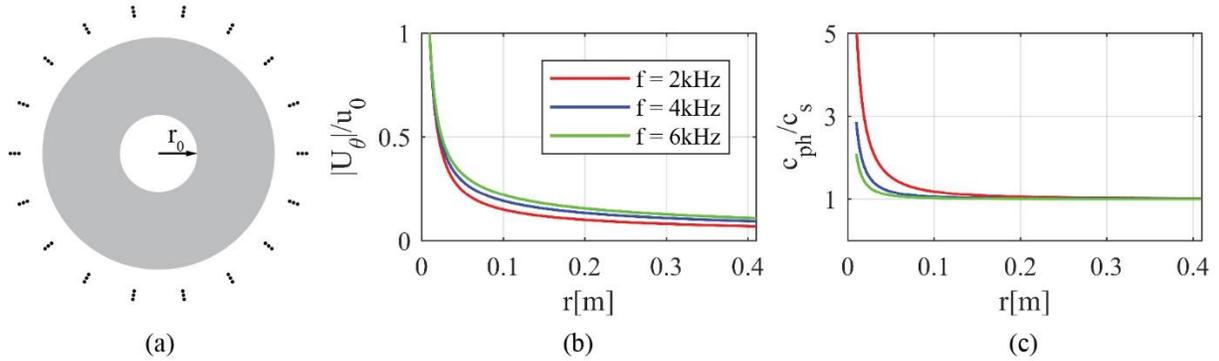

(a)      (b)      (c)

**FIG S2.** (a) Homogenous semi-infinite medium. Radius dependence of (b) normalized harmonic wave amplitudes and (c) normalized phase velocities.

## ZAK PHASE CALCULATION

The Zak phase of the $n^{th}$ mode of EPC1 (Eq. (6)) can be calculated using the method proposed in Xiao et.al. [4],

$$\theta_n^{Zak} = \int_{-\pi/2}^{\pi/2} \left[ i \int_{unitcell} \frac{1}{2P(r)c(r)^2} u_{n,K}^*(r) \partial_K u_{n,K}(r) dr \right] dK \tag{S31}$$

where $u_{n,K}$ is the Bloch periodic eigenfunction, $P(r)$ is equal to $P_i$ inside layer $i$ and $c(r)$ is equal to $\sqrt{M_i/P_i}$ inside layer $i$. We discretize the $k$-space in N points and calculate the Bloch in-cell periodic eigenfunction for each $k$ value applying the transfer matrix method and Bloch theorem to Eq. (6). The origin of the eigenfunctions is chosen to be at the center of layer 1. We then use the discretized version of Eq. (S18) [4] to calculate the Zak phase of each band:

$$\theta_n^{Zak} = -\mathrm{Im}\left[\sum_{i=1}^{N}\ln\left[\int_{unitcell} \frac{1}{2\mathrm{P}(r)c(r)^2} u_{n,K_i}^*(r) u_{n,K_{i+1}}(r)\,dr\right]\right] \quad (S32)$$

Since the dispersion of the EPC2 approximates that of the EPC1 as frequency increases, the same analytical treatment applies at sufficiently high frequencies (well below the Bragg scattering frequency) for EPC2.

## DISCRETE LAYERED EPC

To show feasibility of the EPCs, we run transmission finite element analysis on the EPC2 with $r_0/a = 0.1$ but replacing the continuously varying material properties with different number of discrete layers per unit cell with material properties corresponding to the average of the continuously varying material properties (Fig. S3(a)). We plot transmission of continuously varying EPC2, discrete EPCs and equivalent Cartesian PC in Fig. S3(b). The EPC with 4 discrete layers approximates quite well the continuously varying EPC.

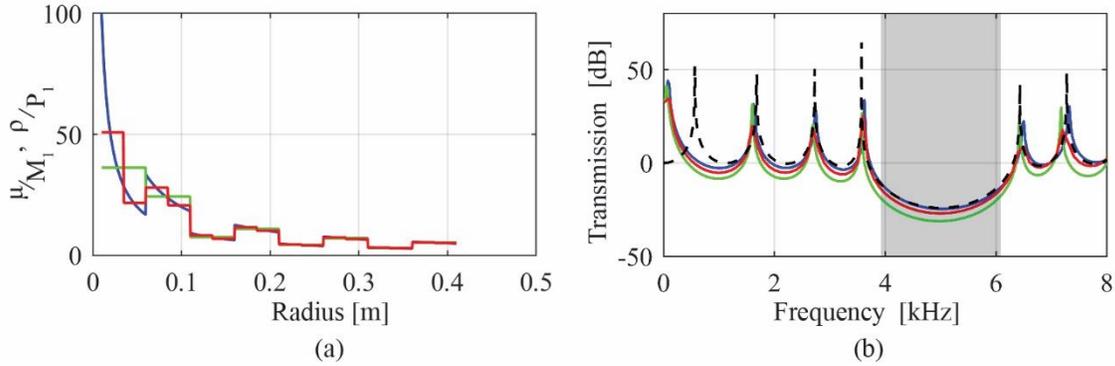

**FIG S3.** (a) Material properties and (b) transmission curves for continuously varying EPC2 (blue), 2 discrete layers per unit cell EPC2 (green), 4 discrete layers per unit cell EPC2 (red). Dashed-black line represent transmission of the equivalent Cartesian system.